\documentclass[twocolumn,showpacs,preprintnumbers,amsmath,amssymb]{revtex4}

\usepackage{graphicx}
\usepackage{dcolumn}
\usepackage{bm}
\usepackage{color}
\usepackage{float}

\begin{document}


\title{Exploring the breakup and transfer coupling effects in $^9$Be elastic scattering}

\author{V. V. Parkar$^{1,2}$\footnote{parkarvivek@gmail.com}}
\author{V. Jha$^1$\footnote{vjha@barc.gov.in}}
\author{S. K. Pandit$^1$}
\author{S. Santra$^1$}
\author{S. Kailas$^1$}

\affiliation{$^1$Nuclear Physics Division, Bhabha Atomic Research Centre, Mumbai - 400085, India}
\affiliation{$^2$Departamento de F\'{i}sica Aplicada, Universidad de Huelva, E-21071 Huelva, Spain}


\begin{abstract}
Two cluster structures of $^9$Be namely, $^5$He+$^4$He and $^8$Be+n have been considered for its breakup to describe the available data on elastic scattering for $^9$Be projectile with $^{28}$Si, $^{64}$Zn and $^{144}$Sm targets. The results of these calculations suggest that the breakup coupling effects are significant for $^5$He+$^4$He cluster model at above the barrier energies, while they are dominant at relatively lower energies for $^8$Be+n model. The addition of one neutron stripping channel in the $^8$Be+n model gives an overall good description of the elastic data for all the systems considered. The couplings generated by breakup in $^8$Be+n model have different behaviour than the coupling effects obtained within the $^5$He+$^4$He model, the former being more prominent at lower energies and for the heavier target systems. The behaviour of extracted dynamic polarisation potentials (DPP) generated due to breakup and one neutron transfer couplings have been investigated.
\end{abstract}

\maketitle

\section{\label{sec:Intro} Introduction}
The study of reaction dynamics for the weakly bound nuclei offer ample opportunities to understand the underlying cluster structure effects. The  measurements of elastic scattering and breakup observable in the experiments involving such nuclei show that the scattering and breakup probabilities are sensitive to the internal cluster structure. The reactions induced by stable weakly bound nuclei $^{6,7}$Li, have been explained quite successfully in terms of their predominant $\alpha$ + d and $\alpha$ + t cluster structures, respectively \cite{Sakuragi83,Sakuragi87,Keeley96,Kumawat,Santra11,Parkar08}. However, the cluster structure of $^9$Be which can be thought in terms of two $\alpha$ particles and a neutron is not clear.  Within the three-body $\alpha$ + $\alpha$ + n picture of $^9$Be, no two constituents alone can form a bound system, a case analogous to that observed in some loosely bound unstable nuclei such as $^6$He and $^{11}$Li, the so called borromean structure. The cluster structure of $^9$Be is not only of interest for studying the reaction dynamics of loosely bound nuclei but is of relevance to certain aspects of nuclear astrophysics. The  nucleosynthesis via the reaction path $\alpha$ + $\alpha$ + $n$ $\rightarrow$ $^9$Be  followed by $^9$Be($\alpha$, $n$)$^{12}$C \cite{Meyer11} is the most efficient path to bridge the stability gaps at A = 5 and A = 8 and knowledge of cluster structure of $^9$Be is vital for calculating the reaction rates. While a three body $\alpha$ + $\alpha$ + n cluster structure picture of  the $^9$Be nucleus is more accurate one, the effective two-body $\alpha$ + $^5$He [$\alpha$ + ($\alpha$ + n)] \cite{Keel01,Keel05} or n + $^8$Be [n + ($\alpha$ + $\alpha$)] \cite{Pandit11,Sign02} cluster configuration can be used successfully in the calculations to explain the reaction dynamics. Experimentally, several measurements of $^9$Be breakup \cite{Brown07,Ful04,Pap07,Raf10} have been performed to quantify the contribution of different cluster decay components in its low-lying excitation spectrum.

There are many experimental evidences which show that the reaction dynamics of weakly bound nuclei are quite different from the tightly bound nuclei. There is a greater importance of the direct reactions, such as breakup or transfer, which may be enhanced owing to the low binding energies and the favourable Q values for selected transfer channels. The coupling effects due to the low-lying resonances and non-resonant continuum arising due to small breakup threshold of a weakly bound nucleus are expected to be quite dominant. In addition, the coupling effect due to transfer channels are also found to be significant. The Continuum Discretized Coupled Channel (CDCC) formalism presents an  effective method to take into account the coupling effects of breakup on elastic scattering and fusion process. In this method, the coupled channels calculations are performed including the bound states of projectile and the continuum of its excitation which is discretized into a finite number of bins. Further, the transfer mechanism can be studied through the Coupled Reaction Channels (CRC) formalism. The coupling effects are manifested in the behaviour of the equivalent Dynamic Polarization Potential (DPP), which may provide a qualitative idea of major couplings that have significant effect on the reaction dynamics \cite{Jha09}. The nature of DPP arising due to the couplings of the breakup and transfer processes is not clear in case of weakly bound nuclei. In general, the breakup couplings are assumed to give repulsive behaviour for the real part of DPP. However, the opposite effect is observed at low energies where the breakup couplings have the dominant dipole contribution. Similarly, while the transfer couplings are expected to give attractive couplings, but in case of positive Q-value reactions, repulsive couplings have been found \cite{Keel05,Keel08}. In addition, it has been shown that the couplings due to breakup also change their nature as a function of energy \cite{Kai10,Jha09}.

In a recent study, we have thoroughly investigated \cite{Pandit11} two cluster configurations of $^9$Be namely, the $^5$He+$^4$He (breakup threshold = 2.46 MeV) and the $^8$Be+n (breakup threshold = 1.67 MeV) through high precision elastic scattering data at sub-Coulomb barrier energies on $^{208}$Pb target. It was demonstrated that, at these sub-Coulomb barrier only $^8$Be+n cluster structure of $^9$Be is able to explain the data satisfactorily. Also, the effect of one neutron transfer was considered in these calculations and found to be dominant even at such sub-Coulomb barrier energies. It is to be noted that, in the previous study by Keeley \textit{et al.} \cite{Keel01}, the cluster structure of $^5$He+$^4$He for $^9$Be was used to explain the elastic scattering angular distribution data of the $^9$Be+ $^{208}$Pb system, at energies around and above the Coulomb barrier. It was also pointed out that the coupling effect of one neutron stripping channel plays an important role. The competing coupling effects of one neutron transfer and the breakup channel for the $^9$Be projectile is crucial for the proper understanding of $^9$Be elastic scattering data for different target systems. Motivated by these investigations, we have attempted to validate the two models of $^9$Be for few other targets ($^{28}$Si \cite{Hugi81}, $^{64}$Zn \cite{Mor00} and $^{144}$Sm \cite{Gomes06}) in different mass regions, where elastic scattering data is available around Coulomb barrier energies.

\section{\label{sec:Caln} Calculations}
To study the effect of breakup and transfer couplings, we have carried out detailed coupled channels calculations. The breakup of the $^9$Be in the reactions for the $^9$Be+$^{28}$Si, $^{64}$Zn and $^{144}$Sm systems has been taken into account by performing the CDCC calculations. In addition to breakup (BU) couplings, the effect of one neutron transfer couplings have also been investigated through the Coupled Reaction Channel (CRC) calculations. The code FRESCO version FRXY.li \cite{Thomp88} is used for  these calculations. The CDCC calculations are performed considering $^9$Be as $^5$He+$^4$He and $^8$Be+n clusters. In the $^5$He+$^4$He cluster picture of $^9$Be, the ground state of $^9$Be (${3/2}^{-}$) is constructed by taking the relative angular momentum L = 0 and 2 between the core $^5$He (${3/2}^{-}$) and the $^4$He (0$^+$) cluster. The L = 2 component is taken in order to account for the reorientation of the highly deformed $^9$Be nucleus. The bound state and the resonances are generated by using the potential between $^5$He (${3/2}^{-}$) and $^4$He (0$^+$) clusters, taken from Ref.\ \cite{Keel01}. The spectroscopic amplitudes 0.81 and 0.5358, obtained from a shell model calculation are used for the two components L = 0 and L = 2 of the g.s. wave function, respectively \cite{Keel01}. Separate g.s. wave functions are obtained for the L = 0 and L = 2 components by adjusting the binding potentials of $^5$He + $^4$He configuration to reproduce the experimental binding energy of $^9$Be. While the deformation in the g.s. is accounted by taking both the L = 0 and L = 2 components explicitly, the continuum states are calculated  as purely L = 0, 1, 2 states. This approximation which omits some terms of the full orthogonal combinations is  seemingly non-trivial. However, in addition to the ground state, the ${5/2}^{-}$ inelastic state at energy 2.43 MeV and the ${7/2}^{-}$ resonance state at 6.38 MeV are generated using the $^5$He+$^4$He cluster model. The breakup calculations including these states along with the $^5$He+$^4$He non-resonant continuum are performed. The $^5$He+$^4$He continuum model space in momentum is limited to 0 $\leq$ k $\leq$ 0.8 fm$^{-1}$. The discretization scheme is suitably modified to take into account the resonances.

In the alternate cluster picture, the $^8$Be+n cluster structure of $^9$Be is assumed. The ground state wave-function of $^9$Be is generated by coupling the valence neutron in 1p$_{3/2}$ state to the $^8$Be (0$^{+}$) core configuration. The Woods-Saxon potential parameters (radius and diffuseness) along with a spin-orbit component for the binding of neutron in $^9$Be is taken from Ref.\ \cite{Lang77}. The ${1/2}^{+}$ and ${5/2}^{+}$ resonance states are generated by using the radius and diffuseness parameters same as that of the ground state while the potential depth is varied. The non-resonant continuum states are generated using the same potential as that of the resonance states. The final CDCC calculations are performed by including the non-resonant continuum and the resonance states. The cluster folding (core-target and valence-target) potentials required in CDCC calculations for constructing $^9$Be + target interaction potential are taken from Refs.\ \cite{England82,Pie10,Obi89,Gomes06,Mori07,Per76,Bal77} as given in Table\ \ref{tab1}. In the final calculations, the depth of real part of optical potential for n+$^{144}$Sm  is normalised by a factor 0.8 at $^9$Be incident energies 37 MeV and below, which is needed to explain the data satisfactorily. The same factor is also used for re-normalizing the depth of real part of optical potential in n+$^{28}$Si at the lowest (12 MeV) energy.

In addition to CDCC calculations for breakup, the CRC calculations for the one neutron stripping channel are simultaneously performed to study its effect on elastic scattering for the $^8$Be+n model. The CDCC wave function calculated as described previously is used in the post-form transfer transition amplitude \cite{neelam12,Moro02} to include coupling of the BU states to the transfer channels. The optical model potentials used in the exit channels are  same as the $^8$Be + target potential parameters as listed in Table\ \ref{tab1} in both the cases. The neutron stripping channel viz; $^{28}$Si($^9$Be,$^8$Be)$^{29}$Si, $^{64}$Zn($^9$Be,$^8$Be)$^{65}$Zn, $^{144}$Sm($^9$Be,$^8$Be)$^{145}$Sm have positive $Q$-values 6.81 MeV, 6.31 MeV and 5.09 MeV respectively. The excited states of the residual nucleus considered in the CRC calculations are chosen where the well defined spectroscopic information is available in the literature \cite{Lang77,Brown10,Brown09} as listed in Table\ \ref{specfac}.

\begin{table*}
\caption{Optical model potentials used in the CDCC and CRC calculations.}
\begin{tabular}
{cccccccccccccc}
\hline
System & V$_{0}$ & r$_{0}$ & a$_{0}$ & W$_{0}$ & r$_{0}$ & a$_{0}$ & V$_{s}$ & r$_{s}$ & a$_{s}$ & W$_{s}$ & r$_{s}$ & a$_{s}$ & Ref. \\
& (MeV) & (fm) & (fm) & (MeV) & (fm) & (fm) & (MeV) & (fm) & (fm) & (MeV) & (fm) & (fm) & \\
\hline \\
$^{4}$He+$^{28}$Si & 50.7 & 1.25 & 0.81 & 20.7 & 1.63 & 0.51 & & & & & & & \cite{Per76} \\
$^{8}$Be+$^{28}$Si & 40.0 &	0.89 & 0.87	& 72.7 & 0.89 & 0.87 & & & & & & &\cite{Bal77} \\
n+$^{28}$Si & 141.5 & 1.29 & 0.60 & & & & & & & 2.1 & 1.29 & 0.60 &\cite{Mori07} \\
$^{4}$He+$^{64}$Zn & 113.6 & 1.57 &	0.46 & 15.0 & 1.67 & 0.18 & & & & & & &\cite{England82} \\
$^{8}$Be+$^{64}$Zn & 126.0 & 1.10 &	0.60 & 17.3 & 1.20 & 0.75 & & & & & & &\cite{Pie10} \\
n+$^{64}$Zn & 70.0 & 1.28 & 0.57 & & & & 2.5 & 1.28 & 0.57 & 6.2 & 1.28 & 0.57 &\cite{Mori07} \\
$^{4}$He+$^{144}$Sm & 50.5 & 1.47 &	0.59 &	18.7 &	1.49 &	0.65 & & & & & & &\cite{Obi89} \\
$^{8}$Be+$^{144}$Sm & 140.0 & 1.06 & 0.71 &	112.0 &	1.06 &	0.71 & & & & & & &\cite{Gomes06} \\
n+$^{144}$Sm & 70.0 & 1.30 & 0.58 & & & & 2.2 & 1.30 & 0.58 & 5.5 & 1.26 & 0.58 &\cite{Mori07} \\
\hline \\
\footnote{For $^{5}$He+target, the same potential of $^{4}$He+target was used with the diffuseness parameter increased by 0.1 fm}
\end{tabular}
\label{tab1}
\end{table*}

\begin{table}[htbp]
\begin{center}
\caption{\label{specfac} Energy levels of residual nuclei and spectroscopic amplitudes (SA) used in the CRC calculations.}\ \\
\begin{tabular}{|ccc|ccc|ccc|}
\hline \multicolumn{3} {|c|} {$^{29}$Si}&
\multicolumn{3} {|c|} {$^{65}$Zn}&
\multicolumn{3} {|c|} {$^{145}$Sm} \\ \hline E& J$^{\pi}$& SA & E& J$^{\pi}$& SA & E& J$^{\pi}$& SA
\\(MeV)&&&(MeV)&&&(MeV)&&\\ \hline
0.00& 1/2$^+$& 0.69 & 0.00& 5/2$^-$& 0.46 & 0.00& 7/2$^-$& 0.78 \\
1.27& 3/2$^+$& 0.83 & 0.06& 1/2$^-$& 0.58 & 0.89& 3/2$^-$& 0.66 \\
2.03& 5/2$^+$& 0.22 & 0.12& 3/2$^-$& 0.41 & 1.11& 13/2$^+$& 0.81 \\
3.62& 7/2$^-$& 0.62 & 0.21& 5/2$^-$& 0.33 & 1.43& 9/2$^-$& 0.92 \\
4.93& 3/2$^-$& 0.84 & 0.86& 1/2$^-$& 0.53 & 1.61& 1/2$^-$& 0.91 \\
5.95& 3/2$^-$& 0.37 & 1.04& 9/2$^+$& 0.79 & 1.66& 5/2$^-$& 0.64 \\
6.38& 1/2$^-$& 0.78 & 1.35& 5/2$^+$& 0.50 & 1.79& 9/2$^-$& 0.58 \\
& & & 1.86& 1/2$^+$& 0.33 & 2.71& 13/2$^+$& 0.55 \\
& & & 4.40& 1/2$^+$& 0.52 & & & \\
& & & 4.78& 1/2$^+$& 0.65 & & & \\
\hline
\end{tabular}
\end{center}
\end{table}
\begin{figure}[htbp]
\includegraphics[width=89mm]{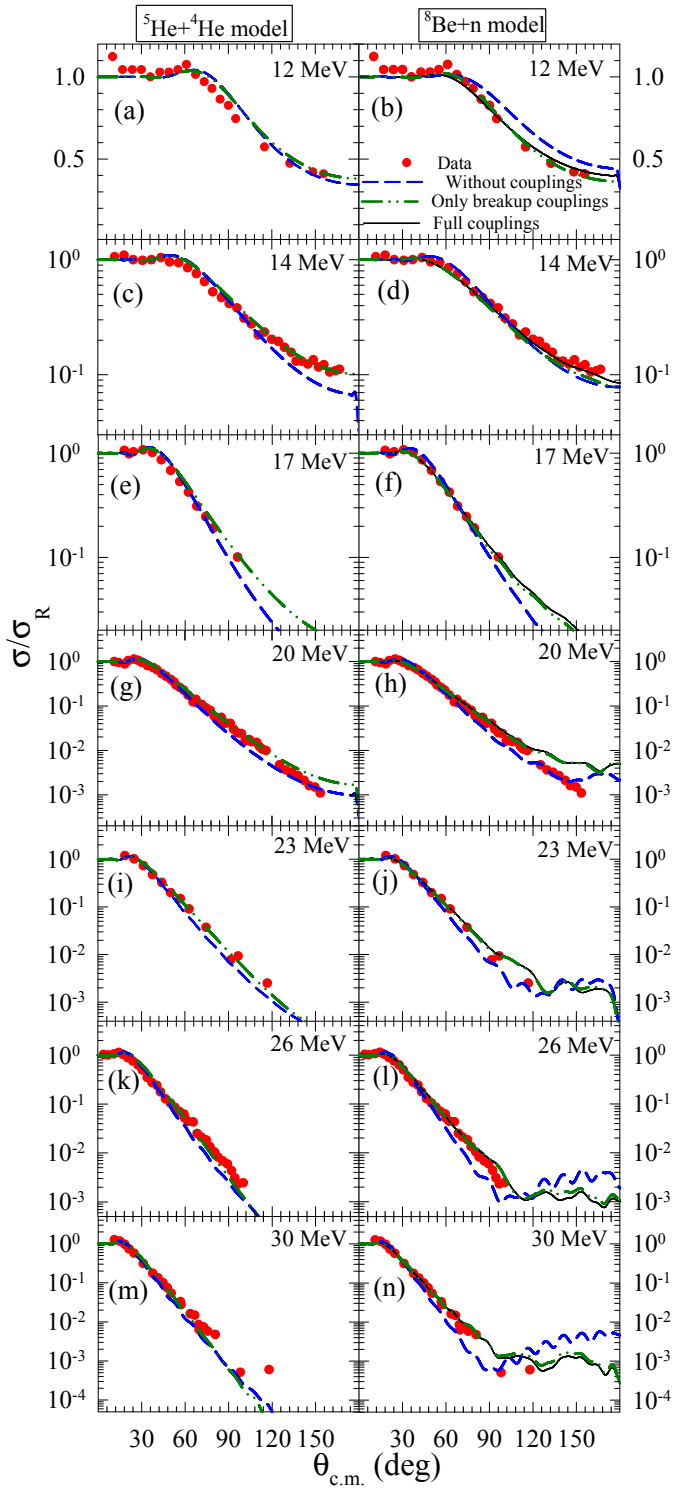}
\caption{\label{elastic_28Si} (Color online) The comparison of measured elastic scattering data for $^9$Be+$^{28}$Si system with the coupled channels calculations from two models: $^5$He+$^4$He (left column) and $^8$Be+n (right column). The dashed, dashed-dot-dot and solid lines are without coupling, only BU couplings, and BU-TR couplings respectively.}
\end{figure}
\begin{figure}[htbp]
\includegraphics[width=89mm]{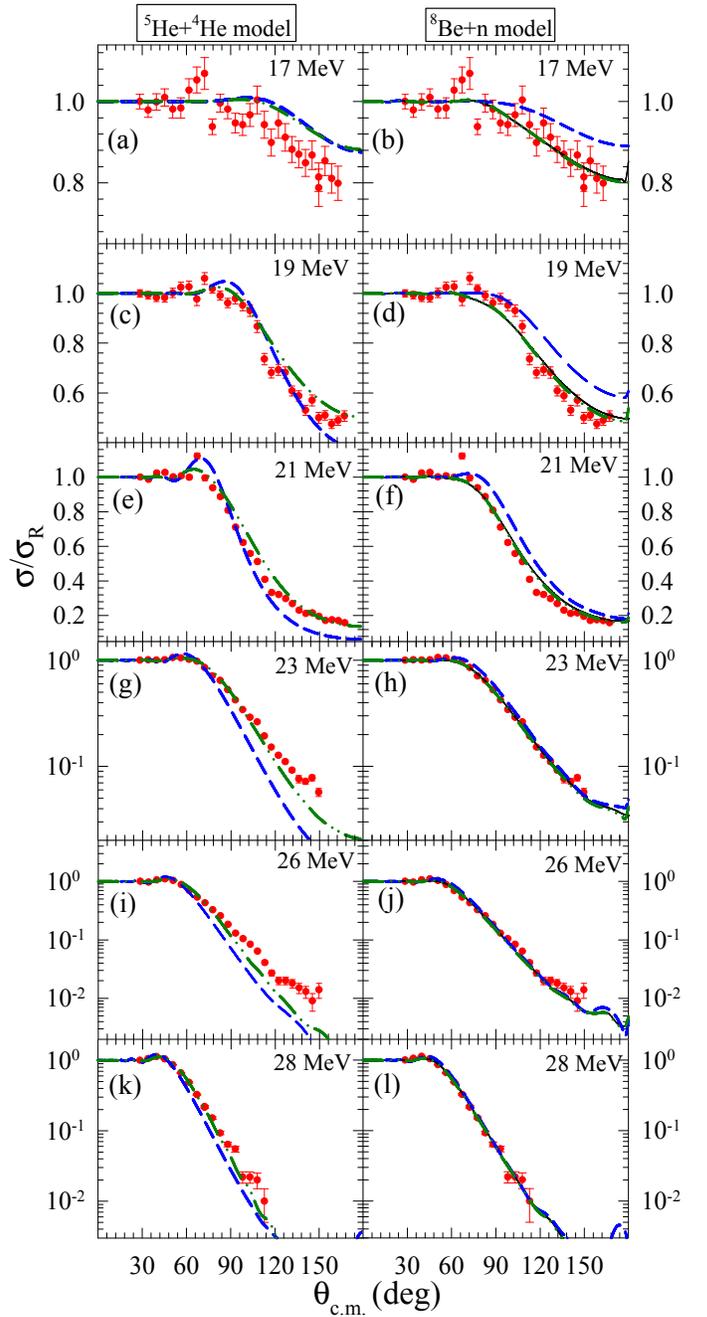}
\caption{\label{elastic_64Zn} (Color online) Same as Fig.\ \ref{elastic_28Si} but for $^9$Be+$^{64}$Zn system.}
\end{figure}
\begin{figure}[htbp]
\includegraphics[width=86mm]{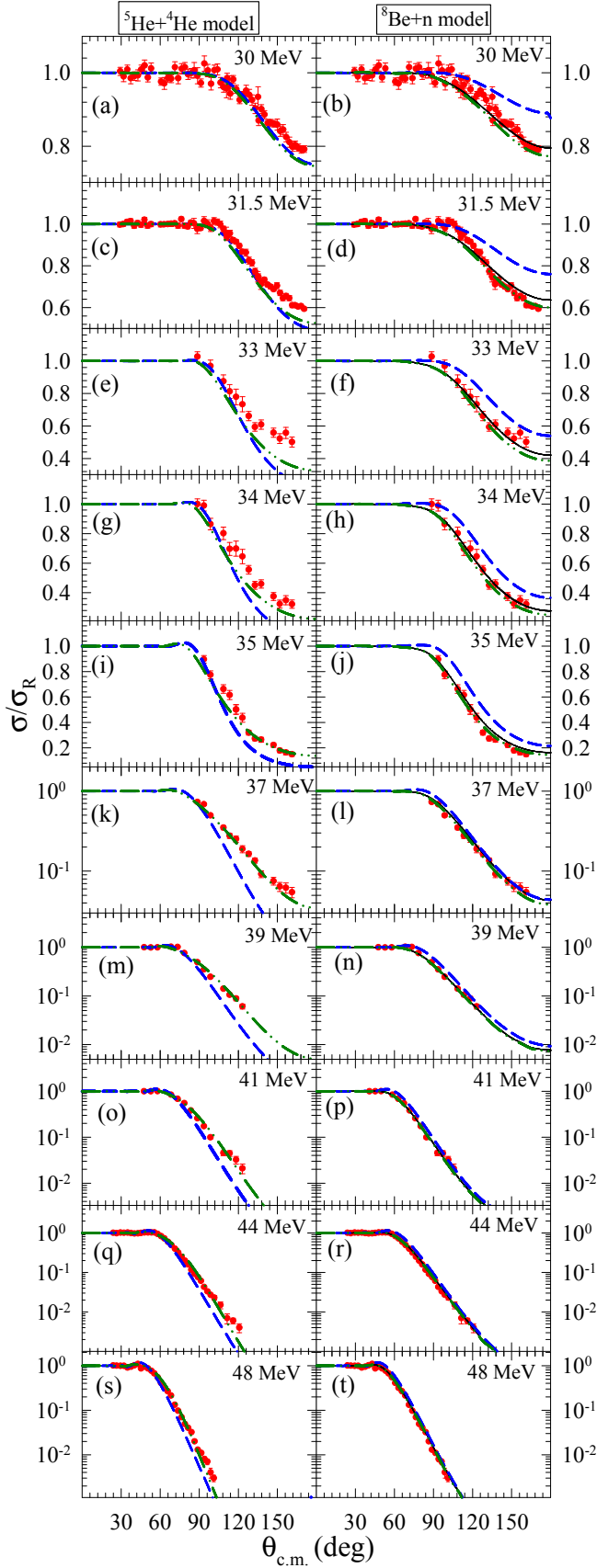}
\caption{\label{elastic_144Sm} (Color online) Same as Fig.\ \ref{elastic_28Si} but for $^9$Be+$^{144}$Sm system.}
\end{figure}
\begin{figure}[htbp]
\includegraphics[width=80mm]{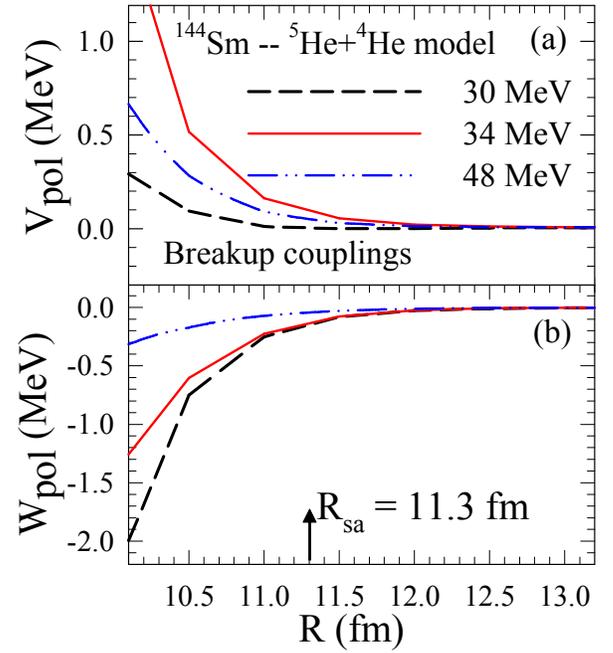}
\caption{\label{Polpot_144Sm_5He} (Color online) Real and imaginary parts of the DPPs around the strong absorption radius (R$_{\textrm{sa}}$=11.3 fm) for $^9$Be+$^{144}$Sm system with the $^5$He+$^4$He model. The potentials are due to the BU couplings.}
\end{figure}
\begin{figure}[htbp]
\includegraphics[width=87mm]{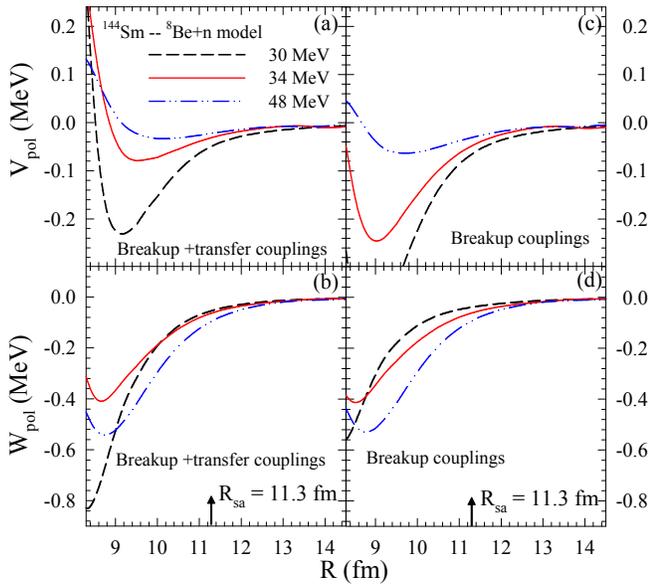}
\caption{\label{Polpot_144Sm_8Be} (Color online) Real and imaginary parts of the DPPs around the strong absorption radius (R$_{\textrm{sa}}$=11.3 fm) for $^9$Be+$^{144}$Sm system with the $^8$Be+n model. The potentials are due to the BU-TR couplings (left column) and BU couplings (right column).}
\end{figure}
\begin{figure}[htbp]
\includegraphics[width=88mm]{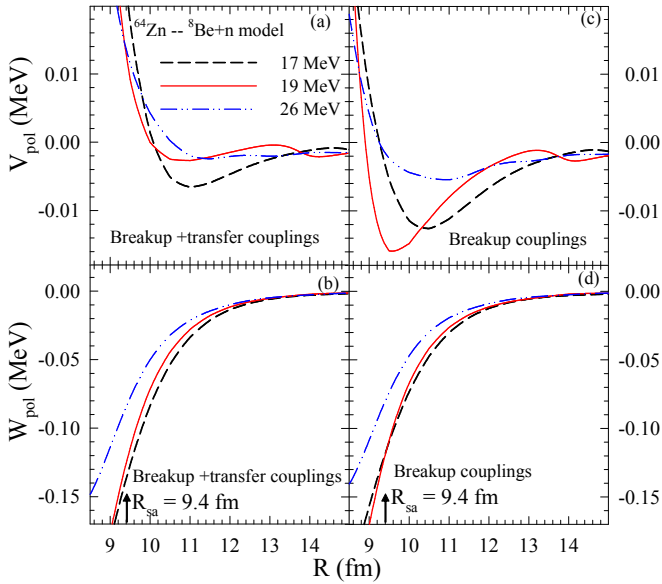}
\caption{\label{Polpot_64Zn_8Be} (Color online) Same as Fig.\ \ref{Polpot_144Sm_8Be} but for $^9$Be+$^{64}$Zn system around R$_{\textrm{sa}}$=9.4 fm with $^8$Be+n model.}
\end{figure}
\begin{figure}[htbp]
\includegraphics[width=88mm]{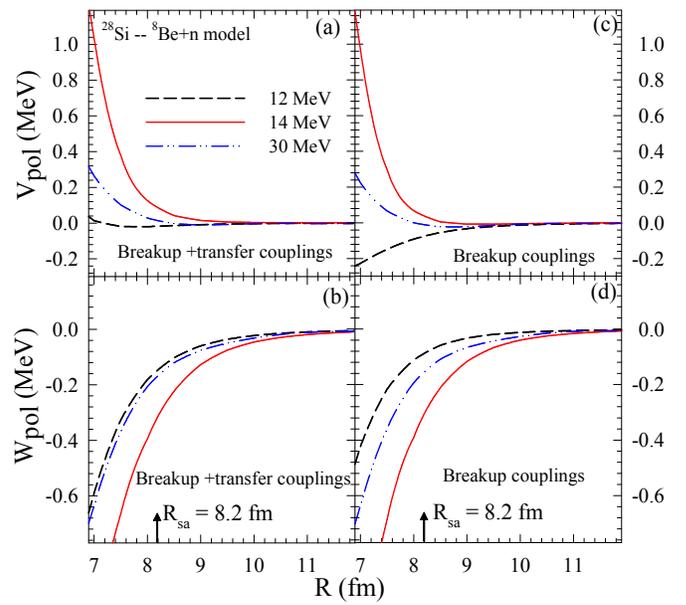}
\caption{\label{Polpot_28Si_8Be} (Color online) Same as Fig.\ \ref{Polpot_144Sm_8Be} but for $^9$Be+$^{28}$Si system around R$_{\textrm{sa}}$=8.2 fm with $^8$Be+n model.}
\end{figure}
\begin{figure*}[htbp]
\includegraphics[width=165mm]{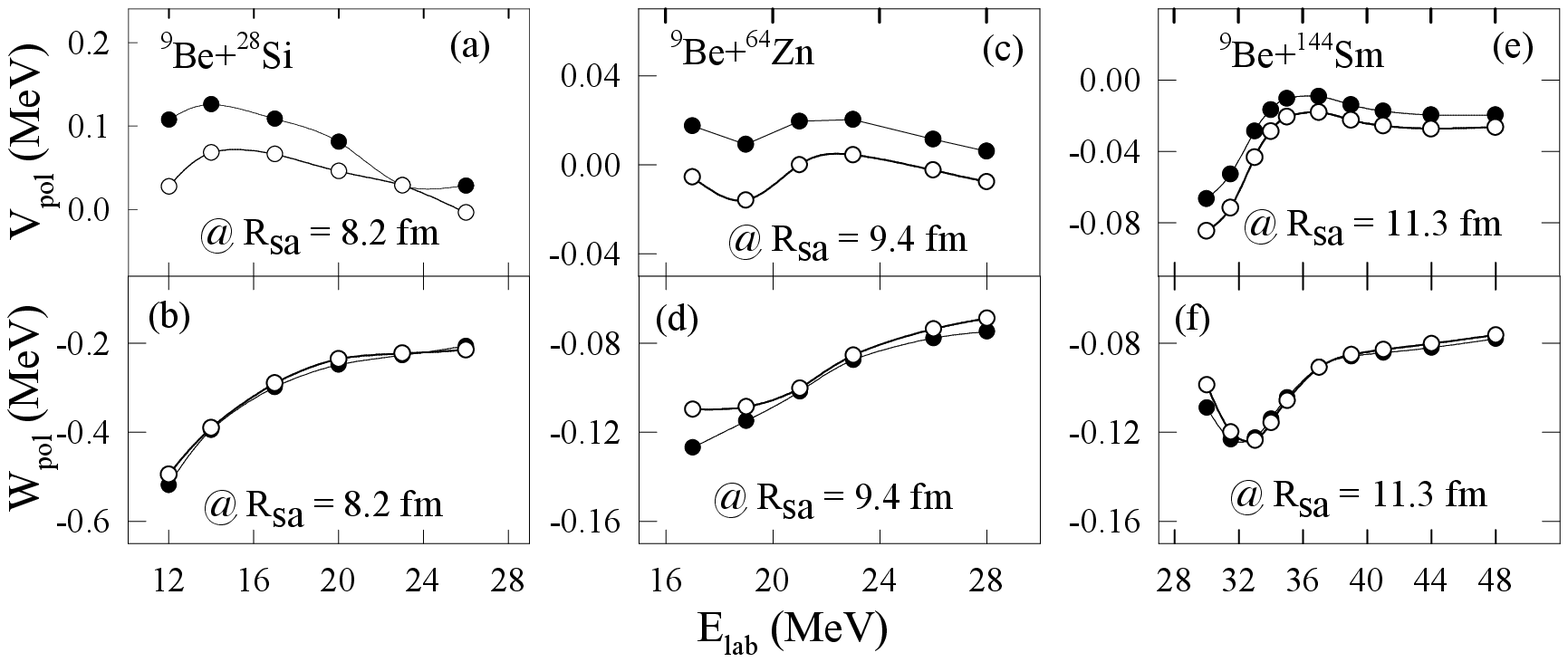}
\caption{\label{thranomaly} Real and imaginary parts of DPPs at the respective R$_{\textrm{sa}}$ due to BU-TR couplings (solid circles) and only BU couplings (empty circles) for $^9$Be+$^{28}$Si, $^9$Be+$^{64}$Zn and $^9$Be+$^{144}$Sm systems with $^8$Be+n model. The solid lines are used to guide an eye.}
\end{figure*}
\section{\label{sec:Result} Results and Discussion}
\subsection{Elastic Scattering}
The measured elastic scattering data along with the calculations for the three systems are shown in Figs.\ \ref{elastic_28Si}, \ref{elastic_64Zn} and \ref{elastic_144Sm} at a range of energies around and above the Coulomb barrier in each case. The calculations without couplings (uncoupled), only BU couplings, and full breakup plus one neutron transfer (BU-TR) couplings are shown with dashed, dashed-dot-dot and solid lines respectively. The effect of the breakup and transfer on the elastic scattering is evident from the difference between the uncoupled calculations and the results obtained from including all BU couplings and subsequently also including the transfer (in the $^8$Be+n model). As can be seen from these figures, at the above Coulomb barrier energies, the $^5$He+$^4$He model gives  significant breakup coupling effects (left column of Figs.\ \ref{elastic_28Si}, \ref{elastic_64Zn} and \ref{elastic_144Sm}). However, at relatively lower energies, more dominant coupling effects are observed in calculations with the $^8$Be+n breakup model (Right column of Figs.\ \ref{elastic_28Si}, \ref{elastic_64Zn} and \ref{elastic_144Sm}). It is observed that, the coupling of breakup channel leads to a rise in the elastic cross sections at the backward angles for the $^5$He+$^4$He model. This is usually the behaviour of breakup continuum couplings for nuclear systems with weakly bound projectiles. In the $^8$Be+n model however, the calculations show a reduction of elastic cross section from the region of Coulomb rainbow to the backward angles. Within this model, the coupling effects of breakup is dominant for a wider range of energies for all the systems. In the test CDCC calculations we have verified that the resonances 2.43 MeV, 5/2$^-$ state in case of  $^5$He+$^4$He model and 1.78 MeV, 1/2$^{+}$ in case of $^8$Be+n model, have the dominant contribution in the BU coupling.

With the inclusion of one neutron transfer channel, the $^8$Be+n model gives an overall good description of the data for these three systems over the entire energy range. However, the transfer couplings are found to affect the elastic cross-section not as much as the BU couplings. This observation is different from the coupling effects observed in $^9$Be+$^{208}$Pb system, where large coupling effects were observed at lower energy (even upto 10 MeV below the Coulomb barrier) due to one neutron transfer. In the present systems, the transfer couplings have a general tendency to give a rise in the backward angle elastic cross section which is opposite to the effect observed in $^9$Be+$^{208}$Pb case \cite{Pandit11}. This can be ascribed to the relatively less positive Q-value for one neutron transfer (+2.27 MeV) which leads to better optimum Q value matching in the latter case.

\subsection{Dynamic Polarisation Potential (DPP)}
To understand the observation from the coupling effects in the calculations for the elastic scattering angular distribution in a better way, we have investigated the behavior of the DPP generated due to these couplings. DPP provides a useful way to simulate the influence of breakup and transfer channel coupling effects by solving the single channel Schrodinger equation for the elastic scattering with an effective potential which comprises of the bare potential and the DPP. The real and the imaginary part of the polarization potentials generated by the BU-TR couplings and only BU couplings are calculated using the prescription of Thompson \textit{et al.} \cite{Thomp88}.

The calculated DPPs due to BU couplings in the $^5$He+$^4$He model for the $^9$Be+$^{144}$Sm system in the vicinity of the strong absorption radii (R$_{\textrm{sa}}$) are shown in Fig.\ \ref{Polpot_144Sm_5He}. It is evident from Fig.\ \ref{Polpot_144Sm_5He} that the BU couplings give rise to repulsive real and attractive imaginary DPPs. Similar behaviour is also observed for the $^9$Be+$^{64}$Zn and $^9$Be+$^{28}$Si systems (not shown here). Next, the calculated DPPs due to BU and BU-TR couplings for the $^8$Be+n model near the respective R$_{\textrm{sa}}$ values for the $^9$Be+$^{144}$Sm, $^9$Be+$^{64}$Zn, and $^9$Be+$^{28}$Si systems are shown in Figs.\ \ref{Polpot_144Sm_8Be}, \ref{Polpot_64Zn_8Be} and \ref{Polpot_28Si_8Be} respectively. The real part of DPPs due to only BU couplings are found to be attractive at larger distances in case of $^9$Be+$^{144}$Sm and $^9$Be+$^{64}$Zn systems. However, in case of $^9$Be+$^{28}$Si system, the real part of DPPs are attractive at the lowest energy (12 MeV), whereas it is repulsive at higher energies.

The attractive nature of the real part of DPP observed in the case of $^9$Be+$^{144}$Sm and $^9$Be+$^{64}$Zn systems is slightly reduced due to the inclusion of transfer couplings (left part of Figs.\ \ref{Polpot_144Sm_8Be}, \ref{Polpot_64Zn_8Be}). For the $^9$Be+$^{28}$Si system, the transfer and BU couplings considered together (left part of Fig.\ \ref{Polpot_28Si_8Be}) lead to an overall repulsive real DPP. The repulsive transfer coupling effect is also observed for the $^9$Be+$^{144}$Sm and $^9$Be+$^{64}$Zn systems, that leads to an overall repulsive real DPP at interior distances. The repulsive coupling effects due to transfer to positive $Q$ value channels has been pointed out also by Keeley \textit{et al.} \cite{Keel05,Keel08}. To summarize the observations on DPP, we encounter a peculiar case where, the real part of DPP is attractive due to BU couplings and it is repulsive due to transfer couplings contrary to the conventional wisdom. This observation underlines the importance of studying the detailed nature of coupling effects for each weakly bound projectile on a case to case basis, before making predictions about the gross features observed in such reactions. For example, phenomena of the fusion suppression and breakup threshold anomaly observed in many weakly bound nuclei have been often ascribed to the repulsive couplings arising due to breakup mode. However, the repulsive couplings due to the transfer processes and attractive couplings due to the breakup mode as observed here may call for a closer look on these effects before making such inferences or making any further predictions.

To gain further insights, the energy dependence of the DPP arising due to the BU and BU-TR couplings at R$_{\textrm{sa}}$ has been studied. The real and imaginary parts of DPPs at the respective R$_{\textrm{sa}}$ due to BU-TR couplings (solid circles) and only BU couplings (empty circles) for $^9$Be+$^{28}$Si, $^9$Be+$^{64}$Zn and $^9$Be+$^{144}$Sm with $^8$Be+n model are shown in Fig.\ \ref{thranomaly}. These calculations have been done with the energy independent potentials listed in Table\ \ref{tab1} without any re-normalization at any energy. From these figures, it is again evident that the inclusion of transfer couplings reduces the strength of real DPPs. The real DPP for the lighter system $^9$Be+$^{28}$Si, is found to be repulsive while for the $^9$Be+$^{144}$Sm system it is attractive at all energies. The attractive DPP in the latter case implies that a usual threshold anomaly could be present for the $^9$Be+$^{144}$Sm system, i.e. at lower energies the real part of the total potential increases in strength (becomes more attractive), while the imaginary part decreases at energies around the Coulomb barrier. However, from the Fig.\ \ref{thranomaly} it is observed that the imaginary part of the potential shows an increase in strength at energies near the Coulomb barrier indicating the presence of unusual threshold anomaly. A similar behaviour in the effective potential has been observed by the optical model analysis of measured data for the $^9$Be+$^{208}$Pb,$^{209}$Bi systems \cite{Yu10}. The additional couplings arising due to transfer lead to an effective DPP, the real part of which is less attractive compared to that given by only BU couplings. While the real part of DPP given by the BU-TR couplings still remains attractive for the relatively heavy target system $^9$Be+$^{144}$Sm, it turns more repulsive for the light target system $^9$Be+$^{28}$Si. The $^9$Be+$^{64}$Zn system is an intermediate case, where the real DPP at R$_{\textrm{sa}}$ is small attractive due to only BU couplings and turns repulsive due to inclusion of transfer couplings. Therefore, it seems that there is a continuous evolution in the behaviour of the real part of DPP from the attractive real for the $^9$Be+$^{144}$Sm system, to the case of light target system $^9$Be+$^{28}$Si, where this is repulsive. Here we would like to remark that the extent of increase of the attractive real potential may be somewhat reduced if the contributions due to continuum couplings due to $^5$He+$^4$He model are also significant.

\section{\label{sec:Sum} Summary}
To summarize, the effect of breakup and transfer couplings have been studied on elastic scattering in the $^9$Be+$^{28}$Si, $^{64}$Zn and $^{144}$Sm systems. The elastic scattering data available around Coulomb barrier have been utilized for these investigations. Two different cluster models of $^9$Be, namely, $^5$He+$^4$He and $^8$Be+n have been investigated. CDCC (breakup) calculations for both the models and CDCC-CRC calculations for the breakup plus one neutron transfer in the $^8$Be+n model have been performed. We obtained a good agreement of the coupled channels calculations with the available data. The $^8$Be+n model gives an overall good description of the data over the entire energy range for all the systems. The calculations for the $^8$Be+n model show significant BU coupling effects at lower energies, whereas for the $^5$He+$^4$He model, large BU coupling effects are seen at energies above the barrier. Detailed investigations using the dynamic polarization potentials show that the attractive real DPP is obtained for the $^8$Be+n breakup model for all the systems at energies below the barrier while the repulsive real DPP is obtained for the breakup via the $^5$He+$^4$He model. In general, larger attractive real DPP due to couplings of the $^8$Be+n breakup are obtained in more heavier systems which remains attractive even at energies above the barrier for these systems. Inclusion of coupling to the single neutron transfer channel in the $^8$Be+n model also gives interesting effects in the present study. In the combined CDCC-CRC calculations, repulsive real DPP are obtained due to transfer processes at all energies, while an overall attractive real DPP is generated due to the dominance of the breakup in $^8$Be+n model at lower energies. However, the imaginary part of the DPP shows an increasing trend towards the lower energy consistent with the phenomenon of unusual threshold anomaly.

An extension of the present work could be the comparison of present results  with those obtained from the calculations using the three body model of $^9$Be in a full four-body CDCC calculation. Such a calculation could provide a unified framework for understanding the effects due to both the $^5$He+$^4$He and $^8$Be+n models apart from studying the possible pure three body effects.

\begin{acknowledgments}
We acknowledge Prof. N. Keeley and the anonymous referee for their valuable comments regarding the calculations. We would like to thank B. Morillon for providing the neutron potentials for all the systems. One of the authors (V.V.P) acknowledges the financial support of INSPIRE Faculty Award, Department of Science and Technology, Govt. of India in carrying out these investigations.
\end{acknowledgments}

\end{document}